\documentstyle{article}
\begin{document}

\begin{center}

{\large \bf  UNIFICATION OF SPINS AND CHARGES IN GRASSMANN SPACE? }

\vspace{1.5cm}

       NORMA MANKO\v C BOR\v STNIK

\vspace{3mm}

{\it Department of Physics, University of Ljubljana, Jadranska 19, }\\
{\it J. Stefan Institute, Jamova 39,61 111 Ljubljana, Slovenia }\\

\vspace{12mm}

{\large ABSTRACT }

\end{center}

\vspace{3mm}

In a space of d $( d > 5) $ ordinary and d Grassmann
coordinates, fields  manifest in an ordinary four-dimensional
subspace as spinor (1/2, 3/2), scalar, vector or tensor
fields with the corresponding charges , according to two kinds of
generators of the Lorentz transformations in the Grassmann
space. Vielbeins and spin connections define gauge
fields-gravitational and Yang-Mills.

\vspace{1mm}
\vspace{1cm}

\begin{large}

{\it 1. Introduction. } In supersymmetric theories $[1]$ fermions
enter into theories through spinor charges, while
electromagnetic, weak and colour charges are defined by
generators of internal (additional) groups. Vierbeins
and spin connections, which define gravitational fields,
transform vectors from freely falling coordinate systems to
external coordinate systems.
Modern Kaluza-Klein theories with fermions try to define all
gauge fields with vielbeins and spin connections  $[2]$,
while they connect charges with Killing vectors which determine
symmetry groups and cause a compactification of extra
dimensions $[2]$ .

In this letter we present the theory in which {\it all internal
degrees of freedom of particles} and {\it fields -spins} and
{\it charges} -are defined by {\it two kinds} of {\it
generators} of the {\it Lorentz
transformations} in the {\it Grassmann} d-dimensional {\it
space} ( forming
the algebra of the Lorentz group $ SO(1,d-1) ) $. The generators of a
spinorial character define spinors: generators with indices
of the four dimensional subspace define spins of spinors, while
the generators with indices of higher dimensions define
charges of these spinors. All gauge
fields - gravitational as well as Yang-Mills including
electrodynamics - are defined by (super)vielbeins, whose spins and
charges  are determined by the generators of the Lorentz group
of a vectorial charater.

In our theory the space has d ordinary $( \{x^a \},commuting) $
and d Grassmann $ ( \{ \theta^a \} , anticommuting,  \theta^a
\theta^b + \theta^b \theta^a =0 ) $ coordinates.

Both kinds of generators of translations in the Grassmann part
of the space have an odd Grassmann character. While those of a
vectorial character form the Grassmann odd Heisenberg algebra,
form those of a spinorial character the Clifford algebra.
Generators of translations and the Lorentz transformations of
both kinds are after the canonical quantization of coordinates
the differential operators in the Grassmann space, and the
Grassmann space ( and accordingly the ordinary space )
has to have  at least $ 5 $ dimensions in order that
the Dirac $ \gamma^a $ operators are the Grassmann even
differential operators .

Since the Lorentz algebra $ SO(2n +1) $ has a regular maximal
subalgebra $ SO(2m) \times SO(2n-2m+1) $  , defines the choice
$ n=7 $ and $ m=5 $, for example,
the Lorentz $ SO(1,4) $ subalgebra, generators of which
determine scalars, spinors, vectors and tensors in the four
dimensional part of the space, while generators of
$ SO(10) $ subalgebra define the electromagnetic, weak and colour
charges ( in the corresponding fundamental and adjoint
representations ). It is the dependence of fields on  Grassmann
coordinates, which determines spins and charges of all fields.

Vector space spanned over a d dimensional Grassmann space has
the dimension $ 2^d $ . Half of vectors have an even, half of
vectors have an odd Grassmann character, demonstrating the
supersymmetry of the theory . The canonical quantization
of fields quantizes the former to bosons, the later to fermions[3].

Generators of translations and the Lorentz transformations in
the ordinary and the Grassmann space form the super-Poincar\' e
algebra[3]. The super-Pauli-Ljubanski vector can be defined as a
generalization of the Pauli-Ljubanski vector with an odd
Grassmann character[3]. It defines spinor charges.

(Super)vielbeins, depending on the ordinary and the Grassmann
coordinates (and connecting (super)vectors of a freely falling
coordinate system to (super)vectors of an external coordinate
system[3] ), define all gauge fields.
A spin connection appearing as a (super)partner of a vielbein
has an odd Grassmann character and describes a fermionic part of
a gravitational field.

\vspace{5mm}

{\it 2. A particle in a freely falling coordinate system. } Since
the space has two kinds of coordinates, commuting ones $ \{x^a\} $
and anticommuting ones $ \{ \theta ^a \} $ , the
geodesics is determined with both kinds of coordinates and with
two parameters: one of a Grassmann even ( $ \tau $ ) another of
a Grassmann odd ( $ \xi $) character  : $ X^a = X^a ( x^a, \theta
^a, \tau, \xi) $. They are called supercoordinates[3,4].
We define the dynamics of a particle by choosing the (simplest )
action

$$ I= \frac{1}{2} \int d \tau d \xi E E^i_A \partial_i X^a E^j_B
\partial_j X^b  \eta_{a b} \eta ^{A B} , \eqno (2.1) $$

where $ \partial _i : = ( \partial _ \tau , {\overrightarrow
{\partial}} _\xi ), \tau^i = (\tau, \xi) $, while $ E^i _A
$ determines a metric on a two dimensional superspace $ \tau ^i
$ , $ E = det( E^i _A )$ . We choose $ \eta _{A A} = 0,
\eta_{1 2} = 1 = \eta_{2 1} $, while $ \eta_{a b} $ is the
Minkowski metric with the diagonal elements $
(1,-1,-1,-1,...,-1) $. The action is invariant under the Lorentz
transformations in the d ( ordinary and Grassmann ) space and
under general coordinate transformations in a two dimansional
space $\tau^i $.

Taking into account that either $ x^a $ or $ \theta^a $ depend on an
ordinary time parameter $ \tau $ and that $ \xi^2 = 0 $ ,
geodesics can be described ( a special choice) as a
polynomial of  $ \xi $ as follows:
$ X^a = x^a + \varepsilon \xi \theta^a $. We choose $ \varepsilon^2 $
to be equal either to $ +i $ or to $ -i $ so that it defines two
possible combinations for supercoordinates. Accordingly we
choose also the metric $ E^i { }_A $[3,5]: $ E^1{ }_1 = 1, E^1{
}_2 = - \varepsilon M, E^2{ }_1 = \xi, E^2{ }_2 = N -
\varepsilon \xi M $, with $ N $ and $ M $ a Grassmann even and
odd parameter, respectivelly. We write $ \dot{A} =
\frac{d}{d\tau}A $, for any $ A $.

In the Grassmann space the left derivatives have to be
distinguished  from the right derivatives, due to the
anticommuting nature of coordinates[5]. We shall make use of
left derivatives only, defined as follows:
$ \frac{\overrightarrow {\partial}}{\partial \theta^a} \theta^b
f= \delta^b_a f - \theta^b
\frac{\overrightarrow{\partial}}{\partial \theta^a} f . $

It turns out[3] that for a particle whose geodesics is defined in
the space of ordinary and Grassmann coordinates, the
Grassmann coordinates are  proportional to their conjugate
momenta $ p^{\theta }_a : = \frac{ \overrightarrow{\partial} L}
{ \partial {\dot{\theta}^a}} = \epsilon^2 \theta^a $. Here $ L$ is the
Lagrange function which follows from the action(2.1) if the
integral over $ d \xi $ is performed[3]. It is appropriatelly to
define generalized coordinates[3]

$$ \tilde{a} ^a := i(p^{\theta a} - i \theta^a) ,\;\;
\tilde{\tilde{a}}{}^a := -(p^{\theta a} + i \theta^a). \eqno
(2.2) $$

The generators of the Lorentz transformations for the
action(2.1), which are

$$ M^{ a b} = L^{a b} + S^{a b} \;,\; L^{a b}= x^a p^b - x^b p^a
\;,\; S^{a b}= \theta^a p^{ \theta b} - \theta^b p^{ \theta a} ,
\eqno (2.3) $$

show that parameters of the Lorentz transformations are the
same in both spaces. The generators may be written  with respect to
operators $ \tilde{a}^a $ and $ \tilde{ \tilde{a}}{ }^a $

$$ S^{a b} = \tilde{S} ^{a b} + \tilde{\tilde{S}}{}^{a b} \;,\;
\tilde{S} ^{a b} = - \frac{i}{4} ( \tilde{a} ^a \tilde{a} ^b -
\tilde{a} ^b \tilde{a} ^a) \;,\; \tilde{\tilde{S}}{}^{a b} = -
\frac{i}{4} (\tilde{\tilde{a}}{}^a \tilde{\tilde{a}}{}^b
-\tilde{\tilde{a}}{}^b \tilde{\tilde{a}}{}^a ) . \eqno      (2.3a) $$

The choice of $ \varepsilon $ makes either $ \tilde{\tilde{a}}{ }^a
(\varepsilon ^2 = -i) $ or $ \tilde{a}^a (\varepsilon ^2 = i) $ equal
to zero and accordingly either $ \tilde{\tilde{S}}{ }^{ab} = 0
\;\;$  or $ \tilde{S}^{ab} = 0 $.

In the canonical procedure the Poisson brackets follow, treating
equivalently the ordinary and the Grassmann space [3]:

$$\{B,A\}_p= -
\frac{ \partial A}{ \partial x^a} \frac{ \partial B}{ \partial
p_a}  + \frac{ \partial A}{ \partial p_a} \frac{ \partial B}{ \partial
x^a} - ( \frac{ \overrightarrow{ \partial A}}{\partial \theta
^a} \frac{ \overrightarrow{ \partial B}}{\partial p^\theta_a} +
  \frac{ \overrightarrow{ \partial A}}{\partial p^\theta_a}
\frac{ \overrightarrow{ \partial B}}{\partial \theta^a})
(-1)^{n_A} , \eqno (2.4) $$

where $n_{A}$ is either one or two depending on whether A has on
odd or an even Grassmann character, respectively.

In the quantization procedure[3] $ -i \{ A,B \}_p $ goes to
either a commutator or to an anticommutator, according to the
Poisson brackets (2.4). The operators $\theta ^a , p^{\theta a}
$ ( in the coordinate representation they become $ \theta^a
\longrightarrow \theta^a , \; p^{\theta}_a \longrightarrow -i
\frac{\overrightarrow{\partial }}{\partial \theta^a} $) fulfil
the Grassmann odd Heisenberg algebra, while the operators

$\tilde{a}^a $and $\tilde{\tilde{a}}{}^{a} $ fulfil the Clifford
algebra:

$$ \{ \theta ^a , p^{\theta b} \} = -i \eta ^{a b} ,\;
\{ \tilde{a}{ }^{a} ,  \tilde{a}{ }^{b} \} = 2\eta^{ab} =
\{\tilde{\tilde{a}}{ }^{a} , \tilde{\tilde{a}}{ }^{b} \} ,
\eqno (2.5) $$

with $ \{ \tilde{a}^a , \tilde{\tilde{a}}{ }^{b} \} = 0 =
\{ \tilde{S}{ }^{ab} , \tilde{\tilde{S}}{ }^{cd} \} $ and
 $\tilde{S}^{ab} = - \frac{i}{4}\lbrack \tilde{a}{ }^{a} ,
\tilde{a}{ }^b \rbrack_{-} , \tilde{\tilde{S}}{ }^{ab} =-\frac{i}{4}
\lbrack \tilde{\tilde{a}}{ }^{a} , \tilde{\tilde{a}}{ }^{b}
\rbrack_{-} $.

Either $L^{ab}$ or $S^{ab}$ or $\tilde{S}^{ab}$ or
$\tilde{\tilde{S}}{}^{ab}$ form the Lie algebra of the Lorentz
group.

It appears[3] that $S^{ab}$ define the adjoint
representations while $\tilde{S}^{ab}$ and $\tilde{\tilde{S}}{}^{ab}$
define the fundamental representations of the Lorentz group.

The constraints which follow from the action(2.1) lead to
the Dirac and to the Klein-Gordon equation

$$ p^a \tilde{a} _a | \tilde{\Psi} > = 0 \;,\; p^a p_a |
\tilde{\Psi}> = 0 , \; with \;  p^a \tilde{a}_a p^b \tilde{a}_b =
p^a p_a . \eqno  (2.6) $$

Since the operator $ \tilde{a}^a $ ( which is a differential
operator in the Grassmann space) has an odd Grassmann
character, it can not be recognized as the Dirac $
\tilde{\gamma}^a $ operator. The dimension of the space d has
therefore to be at least five  $ ( d \ge 5) $  ( which means at
least $5$ odinary and $5$ Grassmann coordinates ) in order that the
generators of the Lorentz transformations $ \tilde{S}^{5 m} , m=
0,1,2,3 $ can be recognized as the Dirac $ \gamma ^m $ operators
of an even Grassmann character

$$\tilde{\gamma} ^m = - \tilde{a} ^5 \tilde{a} ^m = - 2i
\tilde{S} ^{5m} \;,\; m=0,1,2,3.  \eqno  (2.7)$$

It can be checked that in the
four-dimensional subspace $\tilde{\gamma}{ }^{m} $ fulfil the
Clifford algebra $\{\tilde{\gamma}{ }^{m} , \tilde{\gamma}{
}^{n}\}  = \eta{^{mn}} $ , while $ \tilde{S}{ }^{mn} = -\frac{i}{4}
\lbrack \tilde{\gamma}{ }^{m},\tilde{\gamma}{ }^{n}\rbrack_{-}
$.  The operator $ \tilde{\Gamma}=i \tilde{a}{ }^{0} \tilde{a}{
}^{1} \tilde{a}{}^{2} \tilde{a}{ }^{3} = i \tilde{\gamma}{ }^{0}
\tilde{\gamma}{ }^{1} \tilde{\gamma}{ }^{2} \tilde{\gamma}{
}^{3} $ is one of the two Casimir operators of the Lorentz group
$ SO(1,3) $. We have seen, however, that not $ SO(1,3) $ but (
at least )  $ SO(1,4) $  is needed to properly define the Dirac
algebra.

In the case that $ d = 5 $ and $ <\tilde{\psi}|p^{5}|
\tilde{\psi} >= m $  it follows

$$ (\tilde{a}{ }^{m} p_{m} - \tilde{a}{ }^{5} p^5 )|\tilde{\psi} > =
0 = (\tilde{\gamma}{ }^m p_{m} - m )|\tilde{\psi}> \;,\;
m=0,1,2,3. \eqno  (2.8) $$

We show in the references[3] four Dirac four-spinors ( the
polynoms of $ \theta^a $) which fulfil the eqs.(2.8) if $ m \ne
0 $ and four Weyl four-spinors which fulfil the eqs.(2.8) if $
m= 0. $ We define there the super-Pauli-Ljubanski vector which
generates spinor charges. We show also two scalars, two-three
vectors and two four-vectors of an even Grassmann character,
eigenvectors of $ S^{ab} $ (of a vectorial character).

For large enough d not only the generators of the Lorentz
transformations (of a spinorial character ) in the Grassmann
space define ( after the canonical quantization of coordinates )
spinorial degrees of freedom of a particle field in the four
dimensional subspace, they define also quantum numbers of
these fields which may be recognized
as electromagnetic, weak and colour charges. For $ d = 15 $,
for example, we can express the generators of the groups $ U(1),
SU(2) $ and $ SU(3) $ in terms of the generators of the
Lorentz group $ \tilde{S}^{ab} $ with $ a,b = 6,7,...,15 $
while $ SO(5) $ remains to define the spinorial degrees of
freedom in the four dimensional subspace. We find the fundamental
representations of the corresponding Casimir operators as
functions of $ \theta^a $ determining isospin doublets, colour
triplets and electromagnetic singlets. Due to the limited size
of this letter we shall present this results elsewhere.

\vspace{5mm}

{\it 3. A particle in ( Kaluza-Klein ) gauge fields. } We find
the dynamics of a point particle in a gravitational field by
transforming  vectors from a freely falling to an external
coordinate system. To do this, vielbeins  have to be introduced[1,3].
In our case vielbeins ${\bf e}^{ia}{ } _{\mu}, $
depend on ordinary and on Grassmann coordinates, as well as on
two types of parameters $ \tau^i = ( \tau, \xi ) $. Due to two
kinds of derivatives $ \partial_i $ there are two
kinds of vielbeins. The index a
refers to a freely falling coordinate system ( a Lorentz index),
the index $\mu$ refers to an external coordinate system ( an
Einstein index). Vielbeins with the Lorentz index smaller than
five will determine ordinary gravitational fields, those with
the Lorentz index higher than four will define Yang-Mills
fields. Spin connections appear in our theory as ( a part of)
Grassmann odd fields.

We write the transformation of vectors as follows

$$ \partial_i X^a= {\bf e}^{i a} { }_{\mu} \partial_i X^{\mu} \;,\;
\partial_i X^{\mu} = {\bf f}^{i \mu} { }_a \partial_i X^a \;,\;
\partial_i = ( \partial_{\tau} , \partial_{\xi} ) . \eqno   (3.1)$$

 From eq.(3.1) it follows that

$$ {\bf e}^{i a} { }_{\mu} {\bf f}^{i \mu} { }_b = \delta^a { }_b \;,\;
{\bf f}^{i \mu} { }_{a} {\bf e}^{i a} { }_{\nu} = \delta^{\mu} {
}_{\nu} .\eqno (3.2) $$

Again we make a Taylor expansion of vielbeins with respect to
$ \xi $

$$ {\bf e}^{i a} { }_{\mu} = e^{i a} { }_{\mu} + \varepsilon \xi \theta^b
e^{i a} { }_{ \mu b} \;,\; {\bf f}^{i \mu} { }_a = f^{i \mu} {
}_a - \varepsilon \xi \theta^b
f^{i \mu} { }_{a b} \;,\;i=1,2.  \eqno (3.3) $$

Both expansion coefficients  depend again on ordinary
and on Grassmann coordinates. Since  $ e^{ia} { }_{\mu}$ have an even
Grassmann character it will describe the spin 2 part of a
gravitational field. The coefficients $ \varepsilon \theta^{b}
e^{ia} { }_{\mu b}$ have an odd Grassmann
character ($\varepsilon$ is again the complex constant) and are
therefore candidates for spinorial part of a gravitational
field. We shall see that they define spin connections[1,3].

{}From eqs(3.2) and (3.3) it follows that

$$   e^{i a} { }_{\mu} f^{i \mu} { }_b = \delta^a { }_b \;,\;
f^{i \mu} { }_{a}  e^{i a} { }_{\nu} = \delta^{\mu} { }_{\nu}  \;,\;
e^{i a} { }_{\mu b} f^{i \mu} { }_c = e^{i a} { }_{\mu} f^{i
\mu} { }_{c b} \;,\; i=1,2. \eqno (3.2a) $$

We find metric tensor ${\bf g}^{i}_{\mu \nu} = {\bf e}^{ia} {
}_{\mu} {\bf e}^{i}_{a \nu} ,\;
{\bf g}^{i \mu \nu} ={\bf f}^{i \mu} { }_{a} {\bf f}^{i \nu a} ,
i=1,2$,  with an even Grassmann character and the properties
${\bf g}^{i \mu \sigma} {\bf g}^{i}_{\sigma \nu} = \delta ^{\mu}
 { }_{\nu}= g^{i \mu \sigma} g ^{i}_{\sigma \nu}$, with
$g^{i}_{\mu \sigma} = e^{ia} { }_{\mu} e^{i} { }_{a \sigma} $.

We find from eq.(3.1) that vectors in a freely falling and in an
external coordinate system are connected as follows:
$ \dot{x}^a= e^{1 a} { }_{\mu} \dot{x}^{\mu} \;,\; \dot{x}^{\mu} =
f^{1 \mu} { }_a \dot{x}^a\;,\; \theta^a=e^{2 a} { }_{\mu}
\theta^{\mu} \;,\; \theta_{\mu} = f^{2 \mu} { }_a \theta^a, $ and
$ \dot{\theta}^a= e^{1 a} { }_{\mu} \dot{\theta}^{\mu} + \theta^b
e^{1 a} { }_{\mu b} \dot{x}^{\mu} = (e^{2 a} { }_{\mu}
\theta^{\mu})^. =  e^{2 a} { }_{\nu , \mu_x} \dot{x}^\mu
\theta^{\nu} + e^{2 a} { }_{\mu} \dot{\theta}^{\mu} + \dot{\theta}^{\mu}
\overrightarrow{e^{2 a}} { }_{\nu ,\mu_{\theta} } \theta^{\nu}. $

We use the notation $e^{2a} { }_{\nu,\mu_{x}} = \frac{\partial}{
\partial x{^\mu}} e^{2a} { }_{\nu},
\overrightarrow{e^{2a}} { }_{\nu,\mu^{\theta}} =
\frac{\overrightarrow{\partial}}{\partial \theta^{\mu}}
e^{2a} { }_{\nu}$ .

The above equations define the following relations among fields

$$ e^{ 2 a} { }_{\mu b}=0 \;,\;
 \overrightarrow{e^{2 a}} { }_{\nu , \mu_{\theta}} \theta^{\nu} =
e^{1 a} { }_{\mu} - e^{2 a} { }_{\mu} \;,\;
 e^{1 a} { }_{\mu b} = e^{2 a} { }_{\nu , \mu_x} f^{2 \nu} {
}_b, \eqno (3.4)$$

which means that a point particle with a spin sees a spin
connection $ \theta^{b} e^{ia} { }_{\mu b} $ related to a vielbein
$ e^{2a} { }_{\nu}$.

Rewritting the action(2.1) in terms of an external coordinate system
according to eqs.(3.1), using the Taylor expansion of
supercoordinates $ X^{\mu}$ and superfields $ {\bf{e}}^{ia} { }_{\mu}$ and
integrating the action over the Grassmann odd parameter $\xi$
the action

$$ I=\int d\tau \{ \frac{1}{N} g^1_{\mu \nu} \dot{x}^\mu
\dot{x}^\nu - \epsilon^2 \frac{ 2 M}{N} \theta_a e^{1 a} { }_{\mu}
\dot{x}^\mu + \varepsilon^2 \frac{1}{2}( \dot{\theta}^\mu
\theta_a -\theta_a \dot{\theta}^\mu) e^{1 a} { }_{\mu} + \varepsilon^2
\frac{1}{2} (\theta^b \theta_a -\theta_{a} \theta^b ) e^{1
a} { }_{  \mu b} \dot{x}^\mu \} ,$$
$$\eqno  (3.5) $$

defines the two momenta of the system

$$ p_{\mu}=\frac{\partial L}{\partial \dot{x}^\mu}= p_{0 \mu} +
\varepsilon^2 \theta^a \theta^b e^1_{a \mu b} , \;\;
 p^\theta_\mu= \varepsilon^2 \theta_a e^{1 a} { }_{\mu} = \varepsilon^2
(\theta_\mu + \overrightarrow{e}^{2 a} { }_{\nu , \mu_{\theta}}
e^2 { }_{a \alpha} \theta^{\nu} \theta^{\alpha}) . \eqno (3.6)$$

Here $ p_{0 \mu} $ are the covariant ( canonical) momenta of a particle.
For $ p^{\theta}_{a} = p^{\theta}_{\mu} f^{1 \mu} { }_{a}$ it follows
that $ p^{\theta}_{a}$ is proportional to $\theta_{a}$. For a
choice $\varepsilon^{2} = - i $, $ \tilde{a}_{a} = i
(p^{\theta}_{a} - i \theta_{a}),
 $ while $ \tilde{\tilde{a}}_{a}= 0 $. In this case we may write

$$ p_{ 0 \mu} = p_{ \mu} - \frac{1}{2} \tilde{S}^{a b} e^1_{a \mu b}
= p_{ \mu} - \frac{1}{2} \tilde{S}^{a b} \omega_{a b \mu} \;,\;
\omega_{a b \mu}=\frac{1}{2} (e^1_{a \mu b} - e^1_{b \mu a}),
\eqno (3.6a) $$

which is the ussual expression for the covariant momenta in
gauge gravitational fields[1].
One can find  the two constraints

$$ p_0^\mu p_{0 \mu} = 0 = p_{0 \mu} f^{1 \mu} { }_a \tilde{a}^a .
\eqno (3.7)$$

In the quantization procedure the two constraints of eqs.(3.7)
$ p_{0 \mu} f^{1 \mu}{ }_{a} \tilde{a}^{a} p_{0 \nu} f^{1 \nu}
{ }_{b} \tilde{a}^{b} = 0 = p_{0\mu} f^{1\mu} { }_{a} \tilde
{a}^{a}$ have to be symmetrized properly, due to the fact that
fields depend on ordinary and Grassmann coordinates,
in order that the Klein-Gordon and the Dirac equations in the
presence of gravitational fields follow correspondingly.

To see  how  Yang-Mills fields enter into the theory,
the Dirac equation (eq.(3.7), after quantizing
it ) has to be rewritten in terms of fields which determine the
gravitation in the four dimensional subspace and of those fields
which determine the gravitation in higher dimensions. This
should be done by taking  into account the compacification of the space
and looking for states of a particle field and gravitational
fields. This is an ambitious project. In this letter, we
shall limit ourselves on the supposition that a system manifests
itself in the way that only some components of fields are
different from zero and that they depend on Grassmann
coordinates ( which determine spins and charges
of fields ) and on $ x^ \alpha, \; \alpha = 0,1,2,3 $ ( this
should follow from the periodic boundary conditions in a
properly compactified ordinary space, if expectation values of $
p^a = 0 $ , for $ a \ge 6 $ ) while $ m = - p_5 ( e^{15} { }_5 )
{ }^{-1} $ . We find

$$ \tilde{ \gamma}^a f^{1 \mu} { }_a p_{0 \mu} = \tilde{ \gamma}
^m f^{1 \alpha} { }_m ( p_{ \alpha} - \frac{1}{2} \tilde{S} ^{mn}
 \omega_{mn \alpha} + A_{\alpha} ) + m, \eqno (3.8) $$

where $ A_{\alpha} = \tilde{ \tau}^i A^i_{\alpha}, $ with $
\tilde{ \tau}^i A^i_{ \alpha} =
\frac{1}{2} \alpha^{ihk} \tilde{S}^{hk} \beta ^{i
lm} \omega_{lm \alpha} ,h,k,l,m = 6,7,8,..d, $ where $ \alpha
^{ihk} $ and $ \beta^{ihk} $ are two matrices. According to
the subalgebra of the Lie algebra of the Lorentz group $ SO ( d-5
),  \; \tilde{ \tau }^i $ may form the appropriate algebra for the desired
charges. To obtain eq.(3.8) we require that $ e^5{ } _{\mu} = 0
= e^m { }_\sigma, $ with $ m = 0,1,2,3, \sigma = 6,7,8,..d. $

In eq.(3.8) the fields $ \omega _{hk \alpha} $ determine all
Yang-Mills fields, including electromagnetic ones. In the case
that $ e^5 { }_{\alpha} $ is not equal to zero, an additional
term occurs: $ \tilde{ \gamma} ^m f^{\alpha}{ }_m A_{\alpha} $ ,
with $ p_5 f^5 { }_m = f^{\alpha} { }_m A_{\alpha} $, with the
properties of an electromagnetic field[3]. It brings, however,
wrong  magnetic moments of charged particles into the
theory, unless all charged particles are made out of very heavy
constituents, since the mass and the electromagnetic charge of
a particle are related in this case. ( This is the known
unsolved problem of the Kaluza-Klein theory.)

A torsion and a curvature follow from the Poisson brackets
$ \{ p_{0a},  p_{0b} \}_p $, with $ p_{0a} = f^{1\mu} { }_{a} ( p_{\mu} -
\frac{1}{2} \tilde{S}^{cd} \omega_{cd\mu}) $.

We find

$$ \{ p_{0 a} , p_{0 b} \}_p = -\frac{1}{2} S^{c d} R_{ c d a b}
+ p_{0 c} T^c{ }_{a b} , \eqno (3.10) $$
$$ R_{c d a b} = f^{1 \mu} {}_{[a} f^{1 \nu} {}_{b]}
( \omega_{cd\nu,\mu^{x} } + \omega_{c} {}^e{}_{ \mu} \omega_{e d \nu}
+ \overrightarrow{\omega}_{c d \mu , f^{\theta}} \theta^e \omega_e
{ }^f{ }_{ \nu}), $$
$$ T^c { }_{a b} =e^{1c} { }_{ \mu} ( f^{1 \nu} {}_{[b} f^{1 \mu}
{}_{ a]}  {}_{,
\nu} + \omega_{e \nu} { }^d \theta^e f^{1 \nu} {}_{[b}
\overrightarrow{f^{1 \mu}} {}_{  a]} {}_{, d^\theta} ),$$
$ {\rm with } \; A _{[a} B _{b]} = A_a B_b - A_b B_a. $
It has to be pointed out that the Poisson brackets $ \{ p_{0 \mu
}, p_{0 \nu } \}_p $ can be written in terms of the odd Grassmann
fields $ \Psi^a { }_{\mu} = \theta^b \omega_{ab \mu} $ as well

$$ \{ p_{0 \mu} , p_{0 \nu} \}_p = \frac{i}{2} \tilde{a}^a {\cal
D}_ {[ \mu}
\Psi_{a \nu ]},\;\; {\cal D}_{ \mu} \Psi_{a \nu} = \Psi_{a
\nu, \mu} + \frac{i}{2}
\omega_{cd \mu} \tilde{S}^{cd} \Psi_{ a \nu}  ,\eqno(3.10a) $$

where $ {\cal D}_{\mu} $ appears as a {\it covariant derivative of a
spinor- vector field} $  \Psi^a{ }_{\mu} $.

If the action for a free gravitational field is
$ I=\int d^{d} x d^{d}\theta \omega {\cal L}, $
where $\omega$ is a scalar density in the Grassmann space[3],
the Lagrange density $ {\cal L}$
includes  $ det(e^{1a} { }_{\mu})  R $, $ R =  R^{ab}
 {}_{ab} $, or (and) $ det(e^{1a} { }_{\mu}) \;\;  T ^a { }_{cd}
T^{cd} { }_a$ as well as terms with $ {\cal D}_{[ \mu} \Psi_{a
\nu ]} $ combined with $ \tilde{ \gamma}^{\rho} $ into Rarita-
Schwinger like terms[6].

\vspace{3mm}

\vspace{1mm}

{\it 4.Conclusion. } The theory in which the space has d
ordinary and d Grassmann coordinates possesses  supersymmetry
and enables the canonical quantization of coordinates and
fields. In this theory, spins and charges of spinor fields and
gauge fields ( gravitational, Yang-Mills and electromagnetic)
are defined by two kinds of generators of the Lorentz
transformations in the Grassmann space: those of the spinorial
character define properties of spinorial fields, while those of the
vectorial character define properties of gauge fields. The
generators with indices higher than five define charges of
particles and fields, those with indices smaller or equal to five
define spins of particles and fields.
Spin connections have properties of the Rarita-Schwinger field.
All gauge fields -gravitational and Yang-Mills - appear through
vielbeins and spin connections demonstrating their unification.

\vspace{1mm}

{\it 5.Acknowledgement. } The work was supported by Ministry of
Science and Technology of Slovenia and the National Science
Foundation through funds available to the US-Slovenia Joint
Board for Scientific and Technological Cooperation (no.NSF 899).

\vspace{1mm}

\begin{description}
\item[1] J. Wess and J. Bagger,{\it Supersymmetry and Supergravity},
    Princeton Series in Physics ( Princeton University Press, Princeton,
    New Jersey,1983)
\item[2] M.J. Duff, Nucl. Phys. {\bf B 219}, 389 (1983) ,
    M.J. Duff, B.E.W. Nilsson and C.N. Pope, Phys. Lett.
    {\bf B 139}, 154 (1984)
\item[3] N. Manko\v c -Bor\v stnik, Phys.Lett. {\bf B 292}, 25
    (1992),  Il nuovo Cimento {\bf  A 105}, 1461 (1992) ,
    Journ. of Math. Phys. {\bf 34}, 8 (1993), Int. Journal of
    Modern Phys. {\bf A 9}, 1731 ( 1994), IJS.TP.93/15 (submitted
    to Phys. Letters), To appear in
    Proceedings of the Minsk conference Quantum Systems, New
    Trends and Methods, Minsk, May,1994
\item[4] H. Ikemori, Phys.Lett. {\bf B 199}, 239 (1987)
\item[5] F.A. Berezin and M.S. Marinov, {\it The Methods of Second
    Quantization}, Pure and Applied Physics ( Accademic Press, New York,
    1966)
\item[6] D. Luri\'e, {\it Particles and Fields}, Interscience
    Publishers ( John Wiley and Sons, New York 1968)

\end{description}

\end{large}
\end{document}